\documentclass[prl,twocolumn,preprintnumbers,amsmath,amsfonts,amssymb]{revtex4}
\usepackage[letterpaper,breaklinks]{hyperref}
\usepackage[latin1]{inputenc}
\usepackage{bm}
\usepackage{slashed}
\usepackage[T1]{fontenc}
\usepackage{lmodern} 
\usepackage{graphicx}

\renewcommand{\=}{\mathrel{\phantom{=}}}
\newcommand{\s}[1]{\slashed{#1}}
\newcommand{\del}{\partial}

\renewcommand{\L}{\mathcal{L}}
\renewcommand{\O}{\mathcal{O}}

\begin{document}
\title{Gravitational Corrections to Yukawa and \texorpdfstring{$\bm{\varphi^4}$}{phi^4} Interactions}
\date{August 17, 2009}
\author{Andreas Rodigast}
\thanks{\href{mailto:rodigast@physik.hu-berlin.de}{rodigast@physik.hu-berlin.de}}
\author{Theodor Schuster}
\thanks{\href{mailto:theodor.schuster@physik.hu-berlin.de}{theodor.schuster@physik.hu-berlin.de}}
\affiliation{Institut f\"ur Physik, Humboldt Universit\"at zu Berlin, Newtonstra\ss e 15, D-12489 Berlin, Germany}
\preprint{HU-EP-09/36} 
\begin{abstract}
We consider the lowest order quantum gravitational corrections to Yukawa and $\varphi^4$ interactions. 
Our results show that quantum gravity leads to contributions to the running coupling constants if the particles are massive and therefore alters the scaling behavior of the Standard Model. Furthermore, we find that the gravitational contributions to the running of the masses vanish.
\end{abstract}
\maketitle
\section{Introduction}
Einstein's general relativity yields an elegant and successful description of gravity on macroscopic scales, but is---in its perturbatively quantized form---ill-suited as a fundamental theory at arbitrarily high energies because it is non-renormalizable \cite{tHooft:1974bx}. The coupling of the Einstein-Hilbert theory to any type of matter fields leads to non-renormalizable theories as well \cite{deser1974nqe}. Nevertheless, when treated as an effective field theory, as has been established by Donoghue \cite{donoghue1994lqc}, perturbatively quantized Einstein gravity can be used to determine genuine predictions of quantum gravity for energies well below the Planck scale $\smash{M_{\mathrm{Planck}}}\!=\!\smash[b]{G_{\scriptscriptstyle\text{Newton}}^{-1/2}}\!\approx\! \smash[b]{10^{19}}\mathrm{GeV}$. Hence, the effective field theory approach can provide both phenomenologically and methodologically interesting insight into the underlying quantum theory of gravitation, for a review see e.\,g.\,\cite{Burgess:2004rev}.\\
\indent In this context, Robinson and Wilczek \cite{Robinson:2005fj} initiated an intriguing discussion on gravitational corrections to the running of gauge couplings calculated in the framework of effective field theories. They claimed to find gravitational corrections to the running of Abelian and non-Abelian gauge couplings, which would render all gauge theories, including {\small QED}, asymptotically free. However, in a careful reconsideration of the calculations Pietrykowski \cite{Pietrykowski:2006xy} proved, that the back\-ground field method they used yields gauge dependent results. Using a gauge condition independent background field method Toms showed \cite{Toms:2007sk} that the gravitational contributions to the running of gauge couplings vanish. This result has been confirmed by a diagrammatic calculation by one of the authors \cite{ebert2008agc}. Also the inclusion of extra-di\-men\-sional gravity \cite{Ebert:2008ux} did not lead to non-vanishing gravitational corrections.\\
\indent The possibility that quantum gravity alters the scaling behavior of gauge theories has been reopened by a recent work of Toms \cite{Toms:2008dq} showing that there are gravitational corrections to the running of gauge couplings in a system with a non-vanishing cosmological constant. The probably most significant qualitative difference to the models considered before, except the extra-di\-men\-sional scenario, is the introduction of an additional dimensionful parameter, which can be combined with Newton's constant to a dimensionless quantity. This dimensionless quantity naturally appears in front of the logarithmic divergences leading to the running of the coupling. The easiest way to determine these divergences is to use the well established dimensional regularization \cite{tHooft:1972rar}, which in contrast to cut-off regularization respects the diffeomorphism invariance of the model.

One aspect of the gravitational corrections to the Standard Model of particle physics, which has not yet been discussed in the effective field theory approach, are gravitational contributions to the running of the couplings of non-gauge interactions. To close this gap, we consider the lowest order quantum gravitational corrections to the Yukawa and $\varphi^4$ inter\-actions of the Standard Model. This can be achieved by studying a simple model of one scalar and one fermion coupled to gravity. As in the case of a non-vanishing cosmological constant, we need additional dimensionful parameters in order to make logarithmically divergent gravitational contributions possible. In our case these parameters are naturally given by the masses of the matter fields involved in the Yukawa and $\varphi^4$ interactions.

A similar model was recently examined by Zanusso, Zambelli et al.\ \cite{Zanusso:2009bs} using the non-perturbative functional regularization group equation. In their calculation they neglected the renormalization of the field strength; thus, it is not possible to directly compare their and our results.
\section{The setting}
The part of the Standard Model relevant for our considerations can be written in terms of a massive real scalar $\varphi$, representing the Higgs boson, and a massive Dirac fermion $\psi$, representing e.\,g.\ the electron. Both fields are minimally coupled to gravity and can interact via a Yukawa and a $\varphi^4$ interaction
\begin{equation}\label{eq:lagrangian}
\begin{split}
 \L&=\tfrac{2}{\kappa^2}\sqrt{-\bm{g}}\bm{R}+\sqrt{-\bm{g}}\Bigl[\,\overline{\psi}(i\bm{\s{\cal D}}-m_{\scriptscriptstyle\psi})\psi\\
&\=\phantom{\tfrac{2}{\kappa^2}\sqrt{-\bm{g}}\bm{R}+\sqrt{-\bm{g}}(}+\tfrac{1}{2}\bm{g}^{\mu\nu}\partial_\mu\varphi\partial_\nu \varphi-\tfrac{1}{2}m^2_{\scriptscriptstyle \varphi}\varphi^2\\
&\=\phantom{\tfrac{2}{\kappa^2}\sqrt{-\bm{g}}\bm{R}+\sqrt{-\bm{g}}(}-g\, \varphi\overline{\psi}\psi-\tfrac{\lambda}{4!}\varphi^4\Bigr]\,.
\end{split}
\end{equation}
Here $\kappa$ is the gravitational coupling, related to Newton's constant by $\smash{\kappa^2}=32\pi\, G_{\scriptscriptstyle\text{Newton}}$,\, $g$ is the Yukawa coupling, and $\lambda$ is the $\varphi^4$ coupling constant.
The covariant derivative of the fermions is given in terms of the spin connection $\omega^{ab}_{\mu}$ and the vielbeins $e_a^{\phantom{a}\mu}$ by
\begin{equation}
 \bm{\s{\cal D}}=\gamma^a \bm{{\cal D}}_a=\gamma^a e_a^{\phantom{a}\mu}(\partial_\mu+\tfrac{1}{4}\gamma_{bc}\omega_\mu^{bc})\,.
\end{equation}
We expand the metric $ \bm{g}_{\mu\nu}$ around the flat Minkowski background
\begin{equation}
 \bm{g}_{\mu\nu}=\eta_{\mu\nu}+\kappa\, h_{\mu\nu},
\end{equation}
with the  symmetric tensor field $h_{\mu\nu}$ being the graviton.
Note that from now on indices are raised and lowered using the background metric $\eta_{\mu\nu}$. Expanding the Lagrangian in orders of $\kappa$ leads to an infinite series of interactions involving arbitrary numbers of gravitons, e.\,g.\\\
\begin{minipage}[b]{1.5cm}
\centering
 \includegraphics[width=1.3cm]{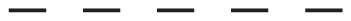}\\[+0.5cm]
$\sim\kappa^0$
\end{minipage}
\begin{minipage}[b]{2cm}
\centering
\includegraphics[width=1.8cm]{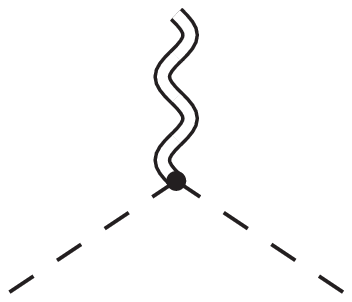}\\
$\sim\kappa^1$
\end{minipage}
\begin{minipage}[b]{2cm}
\centering
 \includegraphics[width=1.8cm]{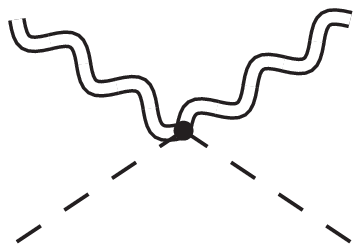}\\
$\sim\kappa^2$
\end{minipage}
\begin{minipage}[b]{2cm}
\centering
\includegraphics[width=1.8cm]{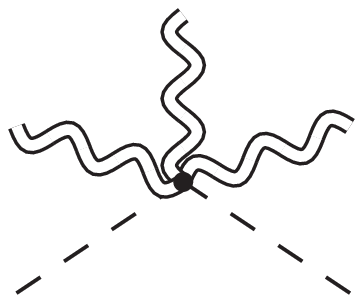}\\
$\sim\kappa^3$
\end{minipage}
\begin{minipage}[b]{0.5cm}
$\cdots$
\end{minipage}
two scalars can couple to any number of gravitons.

General coordinate invariance implies that the Lagrangian \eqref{eq:lagrangian} is invariant under the infinitesimal transformation
\begin{equation}
 \delta_\xi\,h_{\mu\nu}=2h_{\sigma(\mu}\del_{\nu)}\xi^\sigma+\xi^\sigma\del_\sigma h_{\mu\nu}+\tfrac{2}{\kappa}\del_{(\mu}\xi_{\nu)}\;.
\end{equation}
Hence, according to the well-known Faddeev-Popov procedure \cite{Faddeev:1967fc}, we fix this gauge freedom using the harmonic (de Donder) gauge fixing condition
\begin{equation}\label{eq:gfGR}
G_\mu=\del^\nu h_{\mu\nu}-\tfrac{1}{2}\del_\mu h^{\nu}_{\phantom{\nu}\nu}
\end{equation}
and add to the Lagrangian a gauge fixing term as well as the corresponding ghost term
\begin{equation}
 \L_{\text{gauge fixed}}=\L+G_\mu G^\mu-\bar{b}^{\,\mu}\left(\kappa\frac{\delta G_\mu}{\delta \xi^{\nu}}\right)b^\nu\,.
\end{equation}
This leads to a simple form of the graviton propagator
\begin{equation}\label{eq:grav_prop}
 h_{\alpha\beta}\!\raisebox{-0.8ex}{\includegraphics[width=2.5cm]{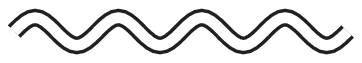}}
 h_{\gamma\delta} \: = \: i\,\frac{\left(\eta_{\alpha(\gamma}\eta_{\delta)\beta} -\tfrac{1}{2}\eta_{\alpha\beta}\eta_{\gamma\delta}\right)}{p^2}\,.
\end{equation}
Since we are only interested in one-loop computations with no external gravitons, the gravitational ghosts are irrelevant in our calculation.
Note that the renormalization of dimensionful parameters, as e.\,g.\ masses, is in general not independent of the chosen gauge. Nevertheless, it is still possible to construct a gauge independent $S$-matrix \cite{antoniadis1986giq}. 
\section{Gravitational Corrections}
What we are interested in is the influence of quantum gravity on the renormalization group flow of our system, which is governed by the Callan-Symanzik equations \cite{Callan:1970yg}. The dependence of the running couplings on the energy scale $\mu$ is determined by their $\beta$ functions, e.\,g.\ 
\begin{equation}
\beta_g = \mu \frac{\mathrm{d} g}{\mathrm{d} \mu}
\end{equation}
for the Yukawa coupling.
 In order to determine the gravitational corrections to the $\beta$ functions of the couplings we have to investigate the divergent part of the proper two-point functions and the proper Yukawa and $\varphi^4$ vertices. These divergences lead to the renormalization of the wavefunctions, masses and couplings. 
\begin{figure}
\centering
 \includegraphics[width=3.5cm]{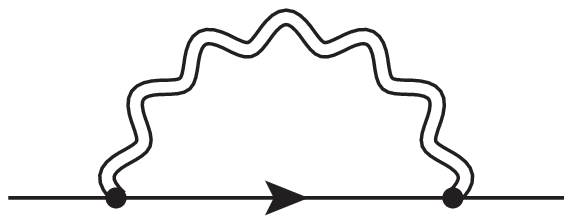}\hspace{1cm}\includegraphics[width=3.5cm]{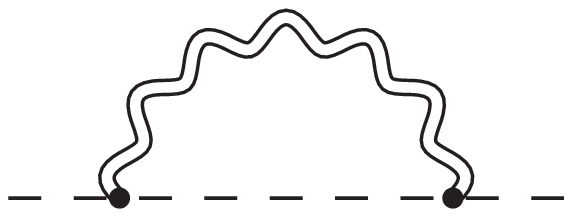}
\caption{Order $\kappa^2$ corrections to the two-point functions.}\label{fig:2point}
\end{figure}
The part of the renormalized Lagrangian relevant for our considerations reads
\begin{equation}
 \begin{split}
 \L_{\text{r}}&=Z_\psi\overline{\psi}i\s{\partial}\psi-Z_{m_{\psi}}m_{\scriptscriptstyle\psi}\overline{\psi}\psi-Z_{\scriptscriptstyle\varphi\overline{\psi}\psi}g\, \varphi\overline{\psi}\psi\\
&\=-Z_{\varphi}\tfrac{1}{2}\varphi\partial^2 \varphi-Z_{m^2_{\varphi}}\tfrac{1}{2}m^2_{\scriptscriptstyle \varphi}\varphi^2-Z_{\varphi^4}\tfrac{\lambda}{4!}\varphi^4+\dots\,,
\end{split}
\end{equation}
here the ellipsis represents the higher order terms involving scalars and fermions as well as all terms involving the graviton.

We use dimensional regularization \cite{tHooft:1972rar}, where the one-loop divergences manifest themselves as poles at $d=4$. Fur\-ther\-more, we apply the Minimal Subtraction scheme where the various $Z$ factors contain solely the divergent pole terms.
Below, we only give the lowest order gravitational corrections $\sim \!\kappa^2$ and omit the well known $\O(\kappa^0)$ terms.
Details of the calculation can be found in \cite{schuster2008dpl}.

The diagrams listed in figure \ref{fig:2point} lead to the following contributions to the wave function renormalizations
\begin{equation}\label{eq:wave_ren}
\begin{split}
 Z_\psi-1&=\frac{\kappa^2}{16\pi^2}\frac{1}{4}m_{\scriptscriptstyle \psi}^2\frac{2}{d-4}\,, \\
  Z_{\varphi}-1&=\frac{\kappa^2}{16\pi^2}m_{\scriptscriptstyle \varphi}^2\frac{2}{d-4}\,,
\end{split}
\end{equation}
which relate the renormalized and bare wavefunctions
\begin{align}
 \psi_0&=Z_\psi^{\frac{1}{2}}\psi\,, &\varphi_0&=Z_\varphi^{\frac{1}{2}}\varphi\,.
\end{align}
The diagrams in figure \ref{fig:2point} also contribute to the mass counterterms
\begin{equation}\label{eq:mass_ren}
 \begin{split}
Z_{m_{\psi}}-1&=\frac{\kappa^2}{16\pi^2}\frac{1}{4}m_{\scriptscriptstyle \psi}^2\frac{2}{d-4}\,, \\
  Z_{m^2_{\varphi}}-1&=\frac{\kappa^2}{16\pi^2}m_{\scriptscriptstyle \varphi}^2\frac{2}{d-4}\,.
 \end{split}
\end{equation}
Together with the wavefunction renormalizations they relate the renormalized and bare masses
\begin{equation}\label{eq:massrenorm}
\begin{aligned}
{m_\psi}_0&=Z_{m_{\psi}}Z_\psi^{-1}m_{\psi}\,,& &\; &{m_\varphi^2}_0&=Z_{m^2_{\varphi}}Z_{\varphi}^{-1}m^2_{\varphi}\,.
\end{aligned}
\end{equation}
Since for both the fermion and the scalar the gravitational contribution to the wavefunction renormalization \eqref{eq:wave_ren} and the mass counterterm \eqref{eq:mass_ren} are equal, there are no gravitational corrections to the running of the masses. We emphasize that this cancellation is unexpected because it is not obvious how the corresponding divergences in each of the diagrams in figure \ref{fig:2point} are linked. Furthermore, the structure of the scalar--graviton and fermion--graviton interactions is very different which makes it remarkable that the cancellation happens two times. 

\begin{figure}
\centering
\begin{minipage}{0.2\textwidth}
\centering
 \includegraphics[width=2.5cm]{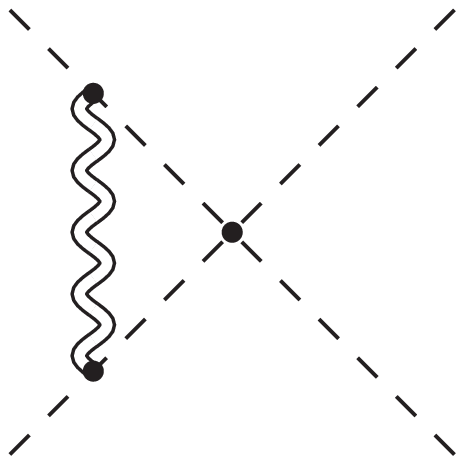}\\
{\footnotesize+ 5 permutations}
\end{minipage}
\begin{minipage}{0.2\textwidth}
\centering
 \includegraphics[width=2.5cm]{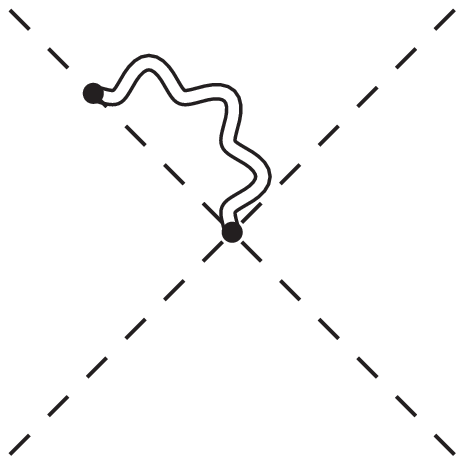}\\
{\footnotesize+ 3 permutations}
\end{minipage}
\caption{Order $\lambda \kappa^2$ corrections to the $\varphi^4$ interaction.}
\label{fig:phi4}
\end{figure}

\begin{figure}
\centering
\includegraphics[width=2.5cm]{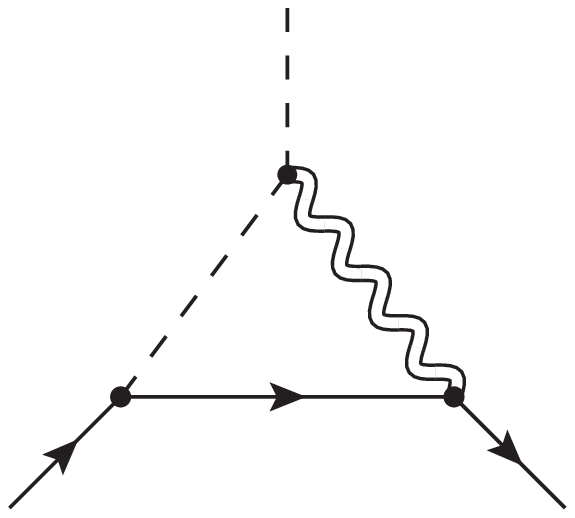}\hspace{0.5cm}\includegraphics[width=2.5cm]{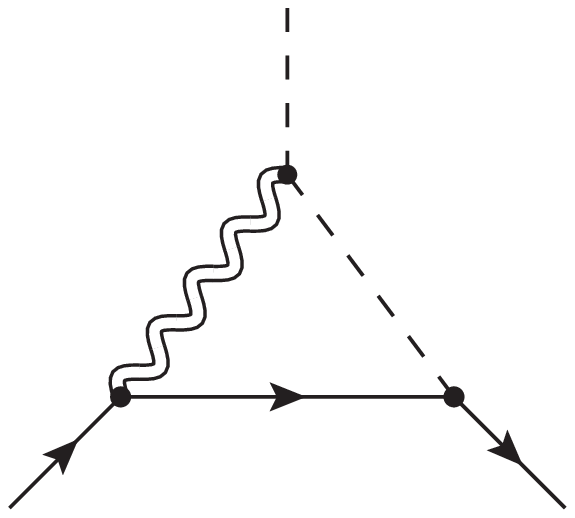}\hspace{0.5cm}\includegraphics[width=2.5cm]{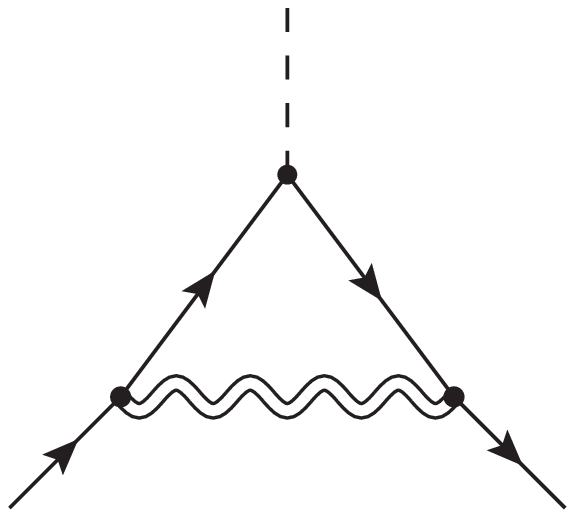}
\includegraphics[width=2.5cm]{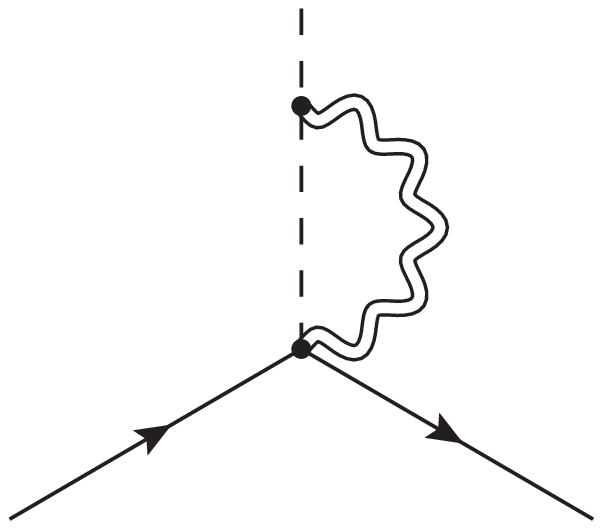}\hspace{0.5cm}\includegraphics[width=2.5cm]{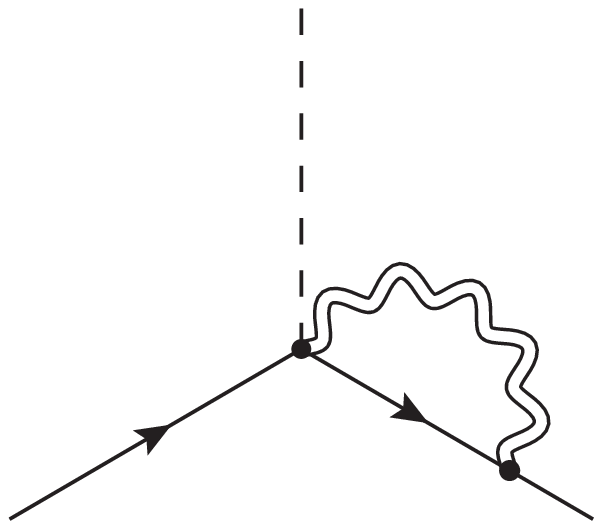}\hspace{0.5cm}\includegraphics[width=2.5cm]{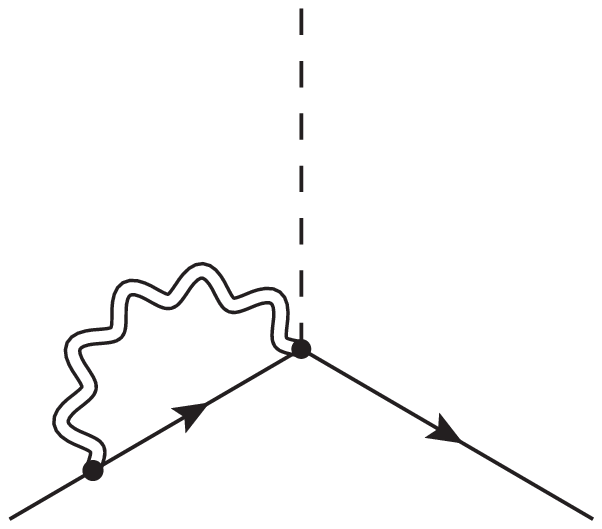}
\caption{Order $g\kappa^2$ corrections to the Yukawa interaction.}
 \label{fig:yukawa}
\end{figure}

To determine the renormalization of the couplings we further need the vertex renormalizations, which can be obtained from the diagrams listed in figures \ref{fig:phi4} and \ref{fig:yukawa} and are given by
\begin{equation}\label{eq:vertex_ren}
 \begin{split}
  Z_{\scriptscriptstyle\varphi\overline{\psi}\psi}-1&=\frac{\kappa^2}{16\pi^2}(\tfrac{3}{4}m_{\scriptscriptstyle \psi}^2+\tfrac{1}{4}m_{\scriptscriptstyle \varphi}^2)\frac{2}{d-4}\,, \\
Z_{\varphi^4}-1&=\frac{\kappa^2}{16\pi^2}4m_{\scriptscriptstyle \varphi}^2\frac{2}{d-4}\,.
 \end{split}
\end{equation}
In dimensional regularization the renormalized and bare couplings are related by
\begin{align}\label{eq:renorm_coupling}
 g_0&=\mu^{\frac{4-d}{2}}Z_{\scriptscriptstyle\varphi\overline{\psi}\psi}Z_\psi^{-1} Z_{\varphi}^{-\frac{1}{2}}g\,, &\;\, \lambda_0=\mu^{4-d}Z_{\varphi^4}Z_{\varphi}^{-2}\lambda\,.
\end{align}
Exploiting that the bare couplings are independent of the renormalization scale $\mu$ it is no problem to compute the gravitational contributions to the $\beta$ functions of the Yukawa coupling
\begin{equation}\label{eq:betafunction_yukawa}
\beta_g=-\frac{\kappa^2}{16\pi^2}(m_{\scriptscriptstyle\psi}^2-\tfrac{1}{2}m_{\scriptscriptstyle\varphi}^2)g
\end{equation}
and the $\varphi^4$ coupling constant
\begin{equation}\label{eq:betafunction_phi4}
 \beta_\lambda=-\frac{\kappa^2}{4\pi^2} m_{\scriptscriptstyle\varphi}^2\lambda\;.
\end{equation}

Note that the $\O(\kappa^0)$ terms in $\beta_g$ and $\beta_\lambda$ as well as the renormalization group equations of $\kappa$, $m_{\scriptscriptstyle\psi}$ and $m_{\scriptscriptstyle\varphi}$ must be taken into account for a complete analysis of the renormalization group flow of the system.

\section{Discussion}
In this Letter we have investigated the gravitational corrections to the Yukawa and $\varphi^4$ interactions of the Standard Model and find that the scaling behavior of the theory is altered by quantum gravity.
For $\sqrt{2}\,m_{\scriptscriptstyle\psi}\!>\!m_{\scriptscriptstyle\varphi}$ both lowest order gravitational corrections are negative. 
Hence, if the non-gravitational as well as higher order gravitational contributions are neglected we find that quantum gravity leads to the asymptotic freedom of the Yukawa and $\varphi^4$ interaction. For the Yukawa coupling this remarkable conclusion should also hold in the region of small couplings and large masses, where the gravitational contributions dominate. By contrast, the $\varphi^4$ coupling cannot vanish in the {\small UV} unless the gravitational coupling is also asymptotically free because of a positive higher order contribution $\sim \kappa^4 m_{\scriptscriptstyle\varphi}^4$ to the $\beta$ function \eqref{eq:betafunction_phi4}.

Including the $\O(\kappa^0)$ terms, the $\beta$ function of the Yukawa coupling has the form
\begin{equation}
 \beta_g=a\,g^3-b\,M^2\kappa^2g\,,
\end{equation}
with $a,b>0$ and $M^2=m_{\scriptscriptstyle\psi}^2-\tfrac{1}{2}m_{\scriptscriptstyle\varphi}^2$. If we assume $M^2>0$ and ignore the running of the masses and the gravitational coupling $\kappa$, we get a UV unstable fixed point at $g^2_\star=M^2\kappa^2\, b/a$. Being proportional to $M/M_{\text{Planck}}$ this fixed point is extremely small and might be without phenomenological relevance.

According to the current state in the search of the Higgs boson,  
the lower bound of the Higgs boson mass is given by $m_{\scriptscriptstyle\varphi}\!\gtrsim\! 114\,\mathrm{GeV}$ \cite{PDG:2008}. Hence, only the gravitational corrections to the Yukawa coupling of the top quark can be negative because it is the only known fermion with a mass greater than $m_{\scriptscriptstyle\psi}>m_{\scriptscriptstyle\varphi}/\sqrt{2}\gtrsim 81\,\mathrm{GeV}$. Taking the known value of the top mass $m_{\text{top}}\!\approx\!171\,\mathrm{GeV}$ \cite{PDG:2008}, we can conclude that gravity only leads to a qualitative change of the scaling of the Yukawa coupling if the Higgs boson is lighter than $\sqrt{2}m_{\text{top}}\approx242\,\mathrm{GeV}>m_{\scriptscriptstyle\varphi}$.

From dimensional grounds it is trivial that the gravitational corrections are absent for \emph{massless} fields since we need a dimensionful parameter in order to have non-vanishing gravitational contributions to the renormalization of the dimensionless couplings. Nevertheless, as has been investigated by Toms \cite{Toms:2008dq} in the case of gauge couplings, there would still be the possibility of contributions arising from a non-vanishing cosmological constant.

The intriguing cancellation between the divergences in the mass counterterms and the wavefunction renormalizations, leading to the vanishing of the gravitational corrections to the running of the masses comes as a nice surprise. At this stage of the investigation it is not clear what causes the cancellation of the divergences. It is not excluded that this cancellation only occurs in de Donder gauge and may be absent in a different gauge, since the renormalization of masses is in general dependent on the chosen gauge \cite{antoniadis1986giq}. If the cancellation turns out to be gauge independent it would be interesting to investigate wether or not this statement is restricted to the lowest order gravitational corrections or holds in general.
\section*{Acknowledgments}
We are grateful to J.\  Plefka for useful comments and his encouragement and want to thank D.\  Ebert for stimulating discussions.
This work was supported by the Volks\-wagen Foundation.
Our computation made use of the symbolic manipulation system \textsc{Form} \cite{Vermaseren:2000nd}.\vspace{2.5ex}

\emph{Note added}---Shortly after this article, Mackay and Toms published their calculation of the gravitational corrections to the mass renormalization of a scalar \cite{Mackay:2009cf}. Keeping the gauge parameters arbitrary, they were able to use the Vil\-ko\-visky-De\-Witt effective action techique and also reproduce our result of a vanishing mass renormalization in the harmonic gauge.

\begingroup\raggedright\endgroup
\end{document}